\documentclass[journal,11pt,onecolumn]{IEEEtran}
\usepackage{epsfig}
\usepackage{subfigure}
\usepackage{graphicx}
\usepackage{times}
\usepackage{cite}
\usepackage{enumerate}
\usepackage{amssymb}
\usepackage{algorithm}
\usepackage{algorithmic}
\usepackage{multirow}
\usepackage{amsmath}
\usepackage{braket}
\newtheorem{corollary}{Corollary}[section]
\newtheorem{theorem}{Theorem}[section]
\newtheorem{lemma}{Lemma}[section]

\newcommand{\geqa}{\ensuremath{\mathrel{\stackrel{\mathrm{(a)}}{\geq}}}}

\begin{document}

\title{On the Stability of Random Multiple Access with Stochastic Energy Harvesting}


\author{
\authorblockN{Jeongho Jeon,~\IEEEmembership{Student Member,~IEEE,} and Anthony Ephremides,~\IEEEmembership{Life Fellow,~IEEE}}
%
\thanks{The material in this paper was presented in part at the IEEE International Symposium on Information Theory, Saint Petersburg, Russia, Aug. 2011.}
\thanks{The authors are with the
Department of Electrical and Computer Engineering and the Institute
for Systems Research, University of Maryland, College Park, MD 20742
USA (e-mail: jeongho@umd.edu; etony@umd.edu).}}
\maketitle

\begin{abstract}

In this paper, we consider random access by nodes that have energy harvesting capability. Each node is equipped with both a queue for storing the arriving packets and a battery for storing the harvested energy chunks, where the packet arrival and the energy harvesting events are all modeled as discrete-time stochastic processes. In each time slot, each node attempts to transmit the head-of-the-line packet in the queue with some probability if its battery is non-empty, and each transmission consumes one chunk of energy. Therefore, the transmission by one node is not just limited by the availability of packets in the queue, but by the availability of energy chunks in the battery. In most of related previous work, it was implicitly assumed that there exists unlimited energy for transmission, which is impractical in many distributed systems such as the ones based on sensors. In this work, we characterize the exact stability region when a pair of bursty nodes, which are harvesting energy from the environment, are randomly accessing a common receiver. The analysis takes into account the compound effects of multi-packet reception capability at the receiver. The contributions in the paper are twofold: first, we accurately assess the effect of limited, but renewable, energy availability due to harvesting on the stability region by comparing against the case of having unlimited energy. Secondly, the impact of the finite capacity batteries on the achieved stability region is also quantified.

\end{abstract}

\begin{keywords}
Stochastic energy harvesting, stability, interacting queues, multipacket reception capacity, random access
\end{keywords}

\section{Introduction}

\PARstart{E}{xploiting} renewable energy resources from the environment, often termed \emph{energy harvesting}, permits unattended operation of infrastructureless distributed wireless networks. There are various forms of energy that can be harvested including thermal, vibration, solar, acoustic, wind, and even ambient radio power \cite{paradiso:energy, chalasani:survey, meninger:vibration}. These aspects of energy harvesting are not examined here. Despite the rapid advancement of hardware technologies, including devices such as solar cells, thermoelectric generators, and piezoelectric actuators, the study of communication systems comprised of nodes that have energy harvesting capability are still in a very early stage. In \cite{ozel:information}, the capacity of the additive white Gaussian noise channel with stochastic energy harvesting at the source node was shown to be equal to the capacity of the channel with an average power constraint equal to the energy harvesting rate. Like most information-theoretic research, however, the result is obtained for a single source-to-destination channel with an infinitely backlogged source and is asymptotic in nature.

When dealing with nodes equipped with non-rechargeable batteries, the common objectives were usually short-term such as maximizing the finite \emph{network lifetime} \cite{kasbekar:lifetime, jeon:neighbor}. The additional functionality of harvesting energy permits our assessment of the system long-term performance such as throughput, fairness and stability. On the other hand, the ALOHA protocol, the simple scheme of attempting transmission randomly, independently, distributively, and based on simple ACK/NACK feedback from the receiver, has gained continued popularity since its creation by Abramson \cite{abramson:aloha}. It is especially suitable for distributed multi-access communication systems due to its simplicity and the independence of the centralized controller. It also serves as a cornerstone benchmark for assessment of performance of more elaborate schemes. In such a context, we revisit the canonical problem of the random access stability when nodes are powered by batteries recharging from randomly time-varying renewable energy sources. To this end, we characterize the stability region of the system which is defined as the set of packet arrival rate vectors for which all queues in the system are stable\footnote{A queue is said to be stable if it reaches a steady state and does not drift to infinity. A formal definition is given in Section \ref{sec:system_model}.}.

In order to put our contribution in perspective, we start with some background on the stability of random access systems. The characterization of the stability region of random access systems for bursty sources (in contrast to \textit{infinitely backlogged} sources, for which the concept does not make sense) is known to be extremely difficult even with unlimited energy for transmissions. This is because each node transmits and thereby interferes with the others only when its queue is non-empty. Such queues are said to be \textit{interacting}, or \textit{coupled}, with each other in the sense that the service process of one depends on the status of the others. Consequently, the individual departure rates of the queues cannot be computed separately without knowing the stationary probability of the joint queue length process, which is  intractable \cite{rao:stability}. This is the reason why most work has focused on small-sized networks and only bounds or approximations are known for the networks with larger number of nodes \cite{tsybakov:ergodicity, rao:stability, Szpankowski:stability, luo:stability, bordenave:asymptotic, naware:stability}. In \cite{tsybakov:ergodicity}, the exact stability region was obtained for the two-node case and for an arbitrary number of nodes with symmetric parameters (that is, equal arrival rates and random access probabilities). In \cite{rao:stability}, a sufficient stability condition for an arbitrary number of nodes with asymmetric parameters was obtained. In \cite{Szpankowski:stability}, the necessary and sufficient stability condition was derived but it can only be evaluated up to the three-node case. The concept of the \emph{instability rank} was introduced in \cite{luo:stability} to further improve the inner bound for the general asymmetric cases. In \cite{bordenave:asymptotic}, an approximate stability region was obtained for an arbitrary number of nodes based on the mean-field asymptotics. All the above results were derived under the collision channel model in which, if more than one nodes transmit simultaneously, none of them are successful. This is too pessimistic assumption today in the sense that a transmission may succeed even in the presence of interference \cite{naware:stability, ghez:stability, tong:multipacket}. In such a context, the two-node stability result was extended to the channel with the multipacket reception (MPR) capability which enables the probabilistic reception of simultaneously transmitted packets \cite{naware:stability}. On the other hand, all existing work on the stability of random access systems implicitly assumes unlimited availability of energy for transmissions, which is impractical for several distributed communication systems.

In this work, we focus attention on the effect of limited availability of energy in each node's battery, that can be recharged by harvesting energy from the environment, on the system stability region when a pair of such nodes are randomly accessing a common receiver with MPR capability. Note that the analysis becomes significantly more challenging than in previous approaches, because the service process of a node depends not only on the status of its own queue and its battery, but also on the status of the other node's queue and battery. The key fact that makes the analysis tractable in this \textit{doubly} interacting system is that the energy consumption is somewhat simplified as it does not depend on the success of the corresponding transmission. For the characterization of the stability region, we first obtain an inner and an outer bound of the stability region for a given transmission probability vector. Since an input rate vector that is outside of the stability region at a given transmission probability vector may be stably supported by another transmission probability vector, determination of the closure of the stability region is necessary and important. Consequently, we take the closure of the inner and the outer bound separately over all feasible transmission probability vectors. The remarkable result is that they turn out to be identical. Therefore, our characterization is exact in the sense that the bounds are tight in terms of the closure. It needs to be pointed out that even without the energy availability constraints, the exact characterization of the stability region in terms of the closure is known only for the two-node case, as also considered here. Finally, we remark that the results presented in this work generalize those of previous work that assumed unlimited energy for transmission over the collision channel \cite{tsybakov:ergodicity, rao:stability} and over the channel with MPR capability \cite{naware:stability}.


The rest of the paper is organized as follows. In Section \ref{sec:system_model}, we revisit the notion of the stability, describe the communication protocol, and explain the channel model and the packet arrival and energy harvesting processes. In Section \ref{sec:main_result}, we present our main result on the stability of the slotted ALOHA with stochastic energy harvesting when nodes are equipped with infinite capacity batteries. The proof of our main result is delayed until Section \ref{section:stability_analysis}. In Section \ref{sec:finite}, we consider the case when nodes are equipped with finite capacity batteries and identify the impact of that restriction on the stability region. Finally, we draw some conclusions in Section \ref{sec:conclusion}.

\section{System Model}\label{sec:system_model}

\subsection{General Description of the Model}

\begin{figure}[t]
\centering
\epsfig{file=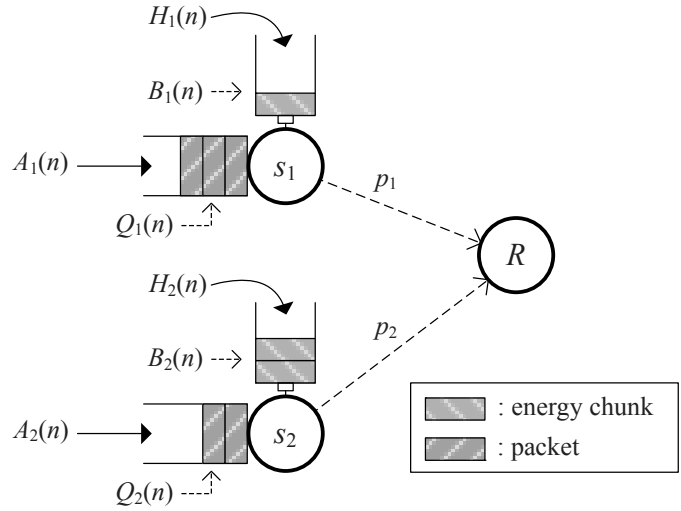,angle=0,width=0.5\textwidth}
\caption{Random access of nodes with stochastic energy harvesting}
\label{fig:sys_model}
\vspace{0cm}
\end{figure}

The system consists of a pair of source nodes randomly accessing a common receiver, each powered by its own battery that can recharge from randomly time-varying renewable energy sources as shown in Fig. \ref{fig:sys_model}. Each node has an infinite size queue for storing the arriving packets, that have fixed length, and a battery\footnote{The battery capacity is first assumed to be infinite and later relaxed to any finite number.} for storing the harvested energy. Time is slotted and the slot duration is equal to one packet transmission time. Energy is harvested in chunks of fixed size and one chunk of energy is consumed in each transmission. That is, the size of the chunk is equal to the slot duration times the power needed to transmit the fixed size packet over the slot duration. Let $(A_i(n), n \geq 0)$ and $(H_i(n), n \geq 0)$ denote the packet arrival and energy harvesting processes at node $i$, respectively. They are modeled as independent and identically distributed (i.i.d.) Bernoulli processes with $E[A_i(n)] = \lambda_i$ and $E[H_i(n)] = \delta_i$. The processes at different nodes are also assumed to be independent of each other. Let $Q_i(n)$ and $B_i(n)$ represent the number of buffered packets and the number of stored energy chunks at node $i$ at the beginning of the $n$-th slot, respectively. Then, $Q_i(n)$ and $B_i(n)$ evolve according to 
\begin{equation}\label{eqn:queue_dynamics}
Q_i(n+1) = Q_i(n) - \mu_i(n) + A_i(n)
\end{equation}
and
\begin{equation}\label{eqn:battery_dynamics}
B_i(n+1) = B_i(n) - 1_i(n) + H_i(n)
\end{equation}
where $\mu_i(n) \in \set{0,1}$ is the actual number of packets that are successfully serviced and, thus, depart from the queue of node $i$ during time slot $n$, and $1_i(n)$ is the indicator function such that $1_i(n)=1$ if node $i$ transmits at time slot $n$, while otherwise $1_i(n)=0$. Node $i$ is said to be \textit{active} if both its data queue and its battery are non-empty at the same time, so that it can then transmit with probability $p_i$. If either the queue or the battery is empty, node $i$ is said to be \textit{idle} and remains silent. Since the nodes are accessing a common receiver, the service variable $\mu_i(n)$ depends not only on the status of its own queue and of its battery but also on the status of the other node's queue and battery. Furthermore, it also depends on the underlying channel model. The collision channel model, for instance, excludes all the possibility of success when packets are transmitted simultaneously.

\subsection{Multipacket Reception Channel}

The channel model with MPR capability considered in this work is a more realistic and general form of a packet erasure model which captures the effect of fading, attenuation and interference at the physical layer, along with the capability of multi-user detectors at the receiver \cite{ghez:stability, tong:multipacket, naware:stability, verdu:multiuser}. Denote with $q_{i | \mathcal{M}}$ the success probability of node $i$ when a set $\mathcal{M}$ of nodes are transmitting simultaneously. The success probability $q_{i | \mathcal{M}}$ is related to the physical layer parameters through
\begin{equation}\label{eqn:sinr_criteria}
q_{i | \mathcal{M}} = \mathrm{Pr} [ \gamma_{i | \mathcal{M}} \geq \theta ]
\end{equation}
where $\gamma_{i | \mathcal{M}}$ denotes node $i$'s signal-to-interference-plus-noise-ratio (SINR) at the receiver given set $\mathcal{M}$ of transmitters and $\theta$ is the threshold for the successful decoding of the received packets, which depends on the modulation scheme, target bit-error-rate, and the number of bits in the packet, i.e., the transmission rate. Once the channel statistics are known, the packet reception probabilities can be readily computed. In Appendix \ref{appendix:A}, the MPR probabilities are obtained in a Rayleigh fading environment. Of course, equation \eqref{eqn:sinr_criteria} is an approximation since it does treat interference as Gaussian noise; however, it is used widely and represents a compromise between accuracy and cross-layer modeling \cite{goldsmith:wireless}.

\subsection{Stability Criteria}

We adopt the notion of stability used in \cite{Szpankowski:stability} in which the stability of a queue is equivalent to the existence of a proper limiting distribution. In other words, a queue is said to be \textit{stable} if
\begin{equation}\label{eqn:definition_stability}
    \lim_{n \rightarrow \infty} \mathrm{Pr}[Q_i(n) < {x} ] = F(x) \ \ \mathrm{and} \ \
    \lim_{ {x} \rightarrow \infty} F(x) = 1
\end{equation}
If a weaker condition holds, namely,
\begin{equation}\label{eqn:definition_substability}
    \lim_{ {x} \rightarrow \infty} \liminf_{n \rightarrow \infty} \mathrm{Pr} [Q_i(n) < {x}] = 1
\end{equation}
the queue is said to be \textit{substable} or bounded in probability. Otherwise, the queue is \textit{unstable}. A stable queue is necessarily substable, but a substable queue is stable if the distribution tends to a limit. If $Q_i(n)$ is an aperiodic and irreducible Markov chain defined on a countable space, which is the case considered in this paper, substability is equivalent to the stability and it can be understood as the recurrence of the chain. Both the positive and null recurrence imply stability because a limiting distribution exists for both cases although the latter may be degenerate. Loynes' theorem, as it relates to stability, plays a central role in our approach \cite{loynes:stability}. It states that if the arrival and service processes of a queue are strictly jointly
stationary and the average arrival rate is less than the average
service rate, the queue is stable. If the average arrival rate is
greater than the average service rate, the queue is unstable and the
value of $Q_i(n)$ approaches infinity almost surely. If they are equal, the queue can be either stable or substable but in our case the distinction is irrelevant as mentioned earlier.


At a given energy harvesting rate and for given transmission probability vectors, the stability region $\mathfrak{S}(\boldsymbol{\delta},\boldsymbol{p})$ is defined as the set of arrival rate vectors $\boldsymbol{\lambda}=(\lambda_1, \lambda_2)$ for which all queues in the system are stable. The stability region of the system $\mathfrak{S}(\boldsymbol{\delta})$ is defined as the closure of $\mathfrak{S}(\boldsymbol{\delta},\boldsymbol{p})$ over all possible transmission probability vectors, i.e., $\mathfrak{S}(\boldsymbol{\delta}) \triangleq \bigcup_{\boldsymbol{p} \in [0,1]^2}
    \mathfrak{S}(\boldsymbol{\delta}, \boldsymbol{p})$.

\section{Main Result}\label{sec:main_result}

\begin{figure}[t]
\centering \subfigure[The case with $\Psi \geq 1$ ($q_{1|\{1,2\}}=0.2, q_{2|\{1,2\}}=0.15$)]{\label{fig:sr_case1}\epsfig{file=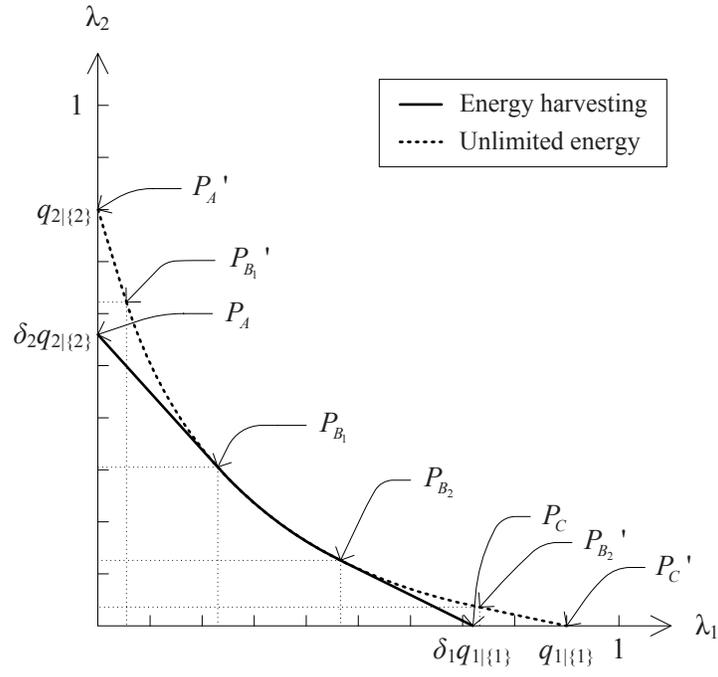,angle=0,width=0.52\textwidth}}\\
\centering \subfigure[The case with $\Psi < 1$ ($q_{1|\{1,2\}}=0.45, q_{2|\{1,2\}}=0.4$)]{\label{fig:sr_case2}\epsfig{file=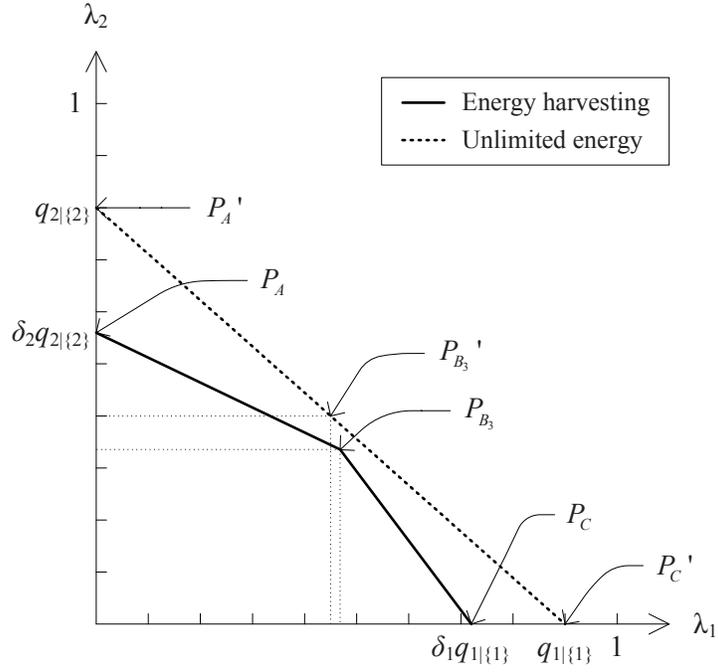,angle=0,width=0.52\textwidth}}
\caption{Two-node stability region $\mathfrak{S}(\boldsymbol{\delta})$ at different MPR probabilities where $\boldsymbol{\delta}=(0.8, 0.7)$ and $q_{1|\{1\}}=0.9, q_{2|\{1\}}=0.8$}
\label{fig:sr_closure}
\end{figure}

This section presents our main results on the stability of slotted ALOHA for the two-node case with stochastic energy harvesting when the capacity of the batteries is assumed to be infinite. Define $\Delta_i = q_{i|\{i\}} - q_{i|\{1,2\}}$, which is the difference between the success probabilities when node $i$ is transmitting alone and when it transmits along with the other node $j$ ($\neq i$). The quantity $\Delta_i$ is strictly positive since interference only reduces the probability of success. 
Let us define the following points in the two-dimensional Euclidean
space to facilitate the description of our main theorem:
\begin{align}
P_A & = \left(0, \delta_2 q_{2|\{2\}} \right) \\
P_{B_1} & = \left(\frac{q_{2|\{2\}} (q_{1|\{1\}} - \Delta_1 \delta_2  )^2}{\Delta_2 q_{1|\{1\}} }, \frac{\Delta_1 \delta_2^2 q_{2|\{2\}} }{q_{1|\{1\}}} \right)\\
P_{B_2} & = \left(\frac{\Delta_2 \delta_1^2 q_{1|\{1\}} }{q_{2|\{2\}}}, \frac{q_{1|\{1\}} (q_{2|\{2\}} - \Delta_2 \delta_1  )^2}{\Delta_1 q_{2|\{2\}}  } \right)\\
P_{B_3} & = \left(\delta_1 ( q_{1|\{1\}} - \Delta_1 \delta_2  ), \delta_2 ( q_{2|\{2\}} -\Delta_2 \delta_1 ) \right) \\
P_C & = \left( \delta_1 q_{1|\{1\}}, 0 \right)
\end{align}
where $P_{B_1}$, $P_{B_2}$, and $P_{B_3}$ are in the first quadrant and $P_A$ and $P_C$ are on $y$ and $x$-axes, respectively. These points can be seen in Fig. \ref{fig:sr_closure}. Let us further define
\begin{equation}\label{eqn:def_psi}
\Psi \triangleq \frac{\Delta_1 \delta_2}{q_{1|\{1\}}} + \frac{\Delta_2 \delta_1}{q_{2|\{2\}}}
\end{equation}
which is non-negative and decreasing as the MPR capability improves.

\begin{theorem}\label{thm:main_theorem}
If $\Psi \geq 1$, the boundary of the stability region $\mathfrak{S}(\boldsymbol{\delta})$ of the slotted ALOHA at a given energy harvesting rate $\boldsymbol{\delta}$ is described by three segments: (i) the straight line connecting $P_A$ and $P_{B_1}$, (ii) the curve
\begin{equation}\label{eqn:main_theorem}
\sqrt{{\Delta_2 \lambda_1}} + \sqrt{{\Delta_1 \lambda_2}} =
\sqrt{q_{1|\{1\}} q_{2|\{2\}}}
\end{equation}
from $P_{B_1}$ to $P_{B_2}$, and (iii) the straight line connecting $P_{B_2}$ and $P_C$. If $\Psi < 1$, it is described by two straight lines: (i) the line connecting $P_A$ and $P_{B_3}$ and (ii) the line connecting $P_{B_3}$ and $P_C$.
%
\begin{proof}
The proof is presented in the next section.
\end{proof}
\end{theorem}
%


In Fig. \ref{fig:sr_closure}, we illustrate the stability region $\mathfrak{S}(\boldsymbol{\delta})$ for different packet reception probabilities. The boundary of the region is indicated by the solid line. The case with unlimited energy, i.e., $\delta_i = 1$, $\forall i \in \set{1,2}$, is also depicted in the figure with the dotted line. The difference between the two regions, therefore, can be understood as the loss due to the limited availability of energy imposed by the variable battery content and the stochastic recharging process.

\begin{corollary}\label{cor:convexity}
If $\Psi > 1$, the stability region $\mathfrak{S}(\boldsymbol{\delta})$ is non-convex, whereas if $\Psi \leq 1$, it is a convex polygon. When $\Psi = 1$, the region becomes a right triangle.
\end{corollary}
This corollary can be easily verified by comparing the slopes of the lines from $P_A$ to $P_{B_1}$ and from $P_{B_2}$ to $P_C$ and those from $P_A$ to $P_{B_3}$ and from $P_{B_3}$ to $P_C$. Specifically, when $\Psi = 1$, the curve \eqref{eqn:main_theorem} shrinks to a point whose coordinates are identical to those of $P_{B_1}$ and $P_{B_2}$ and the slopes of the lines from $P_A$ to $P_{B_1}$ and from $P_{B_2}$ to $P_C$ become identical.

\begin{figure}[t]
\centering
\epsfig{file=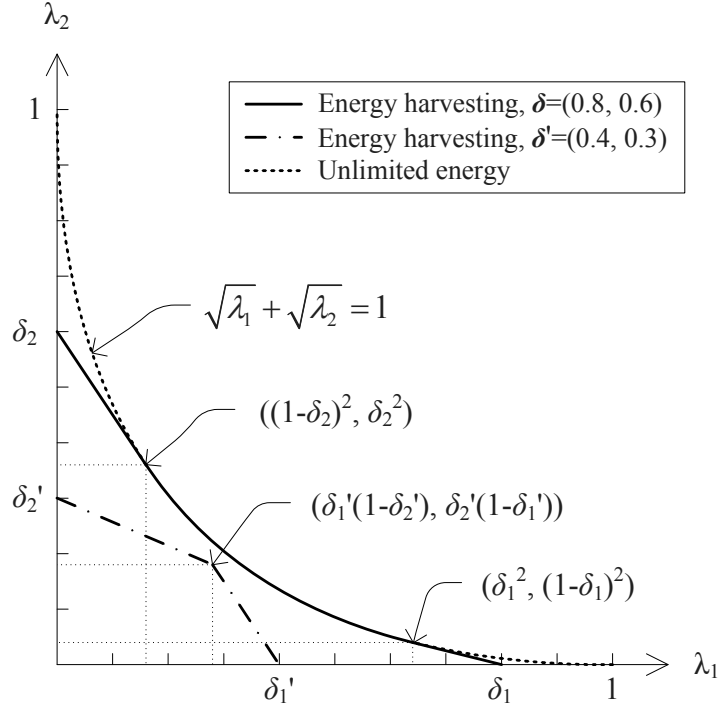,angle=0,width=0.52\textwidth}
\caption{Two-node stability region $\mathfrak{S}(\boldsymbol{\delta})$ under the collision channel model, i.e., $q_{i|\{i\}} = 1$ and $q_{i|\{1,2\}} = 0$, $\forall i \in \set{1,2}$}
\label{fig:stability_region_collision}
\vspace{0cm}
\end{figure}

When we do not have MPR, i.e., when the channel is described by the classical collision channel model, we can obtain the stable region as described in the following corollary and shown in Fig. \ref{fig:stability_region_collision}.
\begin{corollary}\label{cor:stability_collision_channel}
The stability region $\mathfrak{S}(\boldsymbol{\delta})$ of the slotted ALOHA under the collision channel model is described as follows. If $\delta_1 + \delta_2 \geq 1$, its boundary is described by three segments: (i) the line segment connecting $(0, \delta_2)$ and $((1-\delta_2)^2, \delta_2^2)$, (ii) the curve $\sqrt{\lambda_1}+ \sqrt{\lambda_2} = 1$ from $((1 - \delta_2)^2, \delta_2^2)$ to $(\delta_1^2, (1-\delta_1)^2)$, and (iii) the line segment connecting $(\delta_1^2, (1- \delta_1)^2)$ and $(\delta_1, 0)$. If $\delta_1 + \delta_2 < 1$, it is described by two lines: (i) the line segment connecting $(0, \delta_2)$ and $(\delta_1(1 - \delta_2), \delta_2(1- \delta_1))$ and (ii) the line segment connecting $(\delta_1(1 - \delta_2), \delta_2(1- \delta_1))$ and $(\delta_1, 0)$.
\end{corollary}
The corollary is obtained by substituting $q_{i|\{i\}} = 1$ and $q_{i|\{1,2\}} = 0$, $\forall i \in \set{1,2}$, into Theorem \ref{thm:main_theorem}. 

\begin{corollary}
The stability region $\mathfrak{S}(\boldsymbol{1})$ of the slotted ALOHA with the unlimited energy for transmission under the collision channel is the region below the curve $\sqrt{\lambda_1}+ \sqrt{\lambda_2} = 1$ in the first quadrant of the two-dimensional Euclidean space. 
\end{corollary}
This last corollary, which is obtained by substituting $\delta_i = 1$, $\forall i \in \set{1,2}$, into Corollary \ref{cor:stability_collision_channel}, reconfirms the well-known result on the stability of the slotted ALOHA obtained in \cite{rao:stability}.

%

\section{Stability Analysis}\label{section:stability_analysis}

\begin{figure}[t]
\centering
\epsfig{file=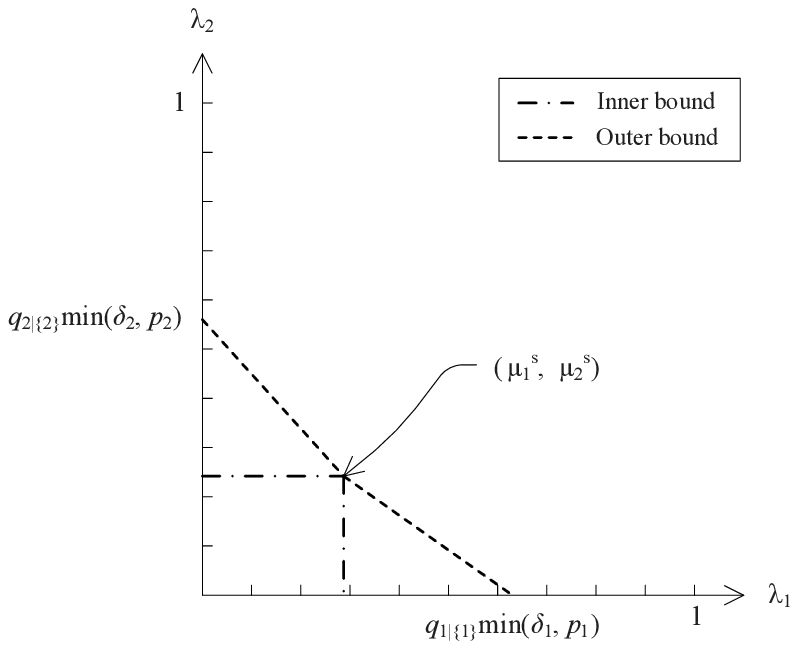,angle=0,width=0.6\textwidth}
\caption{An inner and an outer bound for $\mathfrak{S}(\boldsymbol{\delta},\boldsymbol{p})$}\label{fig:sr_fixed_inner_outer}
\end{figure}

In this section, we prove our main result presented in the previous section. We first derive a sufficient condition for stability in Section \ref{sec:sec:sufficient_condition} and, separately, a necessary condition for stability in Section \ref{sec:sec:necessary_condition} for given energy harvesting rates $\boldsymbol{\delta}$ and transmission probabilities $\boldsymbol{p}$, which yield an inner and an outer bound of $\mathfrak{S}(\boldsymbol{\delta}, \boldsymbol{p})$, respectively, and they are shown in Fig. \ref{fig:sr_fixed_inner_outer}. The achievability and the converse of Theorem \ref{thm:main_theorem} is shown in Section \ref{sec:sec:proof_main_theorem} by taking the closure of the inner and the outer bounds of $\mathfrak{S}(\boldsymbol{\delta}, \boldsymbol{p})$ over $\boldsymbol{p}$ and by observing that these closures turn out to be identical.

\subsection{Sufficient Condition}\label{sec:sec:sufficient_condition}

For the sufficiency, we show that any arrival rate vector $\boldsymbol{\lambda}=(\lambda_1, \lambda_2)$ that is componentwise less than the saturated throughput vector of the system, denoted by $\boldsymbol{\mu}^{s}=(\mu_1^s, \mu_2^s)$, can be stably supported. An input queue is said to be saturated if, after a head-of-the-line (HOL) packet is transmitted from the queue, there is always a packet queued behind it waiting to take the HOL position, i.e., the input buffer is never empty. Since each node transmits with probability $p_i$ whenever its battery is non-empty and each transmission consumes one chunk of energy, the content size of the battery, $B_i(n)$, forms a decoupled discrete-time $M$/$M$/$1$ queue with input rate $\delta_i$ and service rate $p_i$. Consequently, the probability that the battery is non-empty is given by $\min(\frac{\delta_i}{p_i},1)$. The probability of success seen by node $i$ is equal to $p_i q_{i|\{i\}}$ if its battery is non-empty while the battery of node $j(\neq i)$ is empty, and equal to $p_i (1-p_j) q_{i |\{i\}} + p_i p_j q_{i|\{1,2\}}$ if both batteries are non-empty. Hence, the saturated throughput of node $i$ is equal to the average service rate given by
\begin{align}\label{eqn:saturated_throughput_two_queues}
\mu_i^s & = \left\{ p_i q_{i|\{i\}} \left[1-\min\left(\frac{\delta_j}{p_j}, 1 \right) \right] + [p_i (1-p_j) q_{i |\{i\}} + p_i p_j q_{i|\{1,2\}}]\min\left(\frac{\delta_j}{p_j}, 1 \right) \right\} \min\left(\frac{\delta_i}{p_i}, 1 \right) \nonumber \\
& = \min(\delta_i, p_i) (q_{i |\{i\}} - \Delta_i \min(\delta_j, p_j))
\end{align}
for $i,j \in \set{1,2}$ and $i \neq j$. 

\begin{lemma}\label{lemma:stability_sufficient_condition}
The system is stable under the slotted ALOHA if the arrival rate vector $\boldsymbol{\lambda} = (\lambda_1, \lambda_2)$ is componentwise less than or equal to the saturated throughput vector $\boldsymbol{\mu}^s = (\mu_1^s, \mu_2^s)$.

\begin{proof}
It suffices to show that there is no $\boldsymbol{\lambda} \preceq \boldsymbol{\mu}_s$, where `$\preceq$' denotes componentwise inequality, that makes the system unstable. The intuition is that the behavior of a node with unstable queue is statistically identical to that with saturated queue as time goes to infinity. This is because a queue being unstable, or equivalently transient, implies that its size grows to $\infty$ without emptying with a nonzero probability. Therefore, if one queue, say queue $i$, becomes unstable at some $\boldsymbol{\lambda}'$, then the corresponding input rate component $\lambda_i'$ is greater than the service rate seen in the saturated throughput $\mu_i^s$. 

We formally support this argument as follows. Let $\epsilon_i$ be any real number satisfying $0 \leq \epsilon_i <\mu_i^s$ and let $\lambda_i = \mu_i^s - \epsilon_i$, $\forall i \in \set{1,2}$, such that $\boldsymbol{\lambda} \preceq \boldsymbol{\mu}^s$ and suppose that the system is unstable with the chosen input rate vector. The instability of a system implies that at least one queue in the system is unstable. Let us first suppose that both queues are unstable such that their queue sizes, $Q_i(n)$, $\forall i \in \set{1,2}$, grow to infinity without emptying with nonzero probability. Since the number of recurrences of the empty-state is finite with probability 1, the probability that each battery is non-empty approaches $\min(\frac{\delta_i}{p_i},1)$ as $n \rightarrow \infty$, for $i \in \set{1,2}$. Then, the limiting expectation on the actual rate serviced out of  queue $i$, denoted by $\lim_{n \rightarrow \infty} E[\mu_i(n)]$, is equal to $\mu_i^{s}$ in \eqref{eqn:saturated_throughput_two_queues}. On the other hand, from the queueing dynamics \eqref{eqn:queue_dynamics}, for any $n>0$, we have
\begin{equation}
Q_i(n) - Q_i(0) = \sum_{k=0}^{n-1} A_i(k) - \sum_{k=0}^{n-1} \mu_i(k)
\end{equation}
By taking expectations, diving by $n$, and taking a limit as $n \rightarrow \infty$ we have
\begin{equation}\label{eqn:limit_queue_dynamics}
\lim_{n \rightarrow \infty}\frac{E[Q_i(n)]}{n} = \lambda_i - \lim_{n \rightarrow \infty} \frac{1}{n} \sum_{k=0}^{n-1} E[\mu_i(k)]
\end{equation}
where we use the fact that the effect of the initial queue size disappears as $n \rightarrow \infty$ because of the stationarity of the arrival process. The instability of queue $i$ implies that the left-hand-side (LHS) of \eqref{eqn:limit_queue_dynamics} is strictly positive and, then, it must be the case that 
\begin{equation}
\lambda_i = \mu_i^s - \epsilon_i > \lim_{n \rightarrow \infty} E[\mu_i(n)] =  \mu_i^s
\end{equation}
which is impossible because $\epsilon_i$ is non-negative.

Consider now the case when only one of the queue in the system, say node $i$, is unstable and the other queue at node $j (\neq i)$ is stable, i.e., the limiting distribution exists only for the queue at node $j$ while the queue at node $i$ is transient. At the steady-state of queue $j$, the probabilities that the queue and the battery at node $i$ are non-empty are 1 and $\min(\frac{\delta_i}{p_i},1)$, respectively. These do not depend on $j$. In other words, node $i$ randomly interferes with node $j$ with some probability which only has the effect of lowering the average success probability of node $j$. Since queue $j$ is stable, \textit{the average input rate is equal to the average output rate}, which can be deduced by setting the LHS of \eqref{eqn:limit_queue_dynamics} to zero. Additional care should be taken here, because, not all the random variables with a well-defined distribution function have finite expectation, when the distribution is heavy-tailed with tail exponent that is less than 1 \cite{cappe:long-range}. However, in most standard queueing systems including $M$/$M$/1, $M$/$G$/$1$, and $G$/$M$/1 systems, the queue size exhibits exponential tail \cite{kleinrock:queueing, ross:bounding}. The queue $j$ under consideration is a decoupled discrete-time system that is a variant of $M$/$M$/1 system whose service is paused when the battery is empty. Since the inter-arrival times of energy chunks follow geometric distribution, and since the energy process is independent of the data process and the channel, the limiting distribution of queue $j$ is also not heavy-tailed. Let us now compute $\mu_{j|\mathrm{active}}$ which is defined as the expected number of packets that is successfully serviced from node $j$ given that node $j$ is active, i.e., given that both its data queue and its battery are non-empty. Since the  probability of success seen by node $j$ is $p_j q_{j|\{j\}}$, if node $i$'s battery is empty, and $p_j (1-p_i) q_{j |\{j\}} + p_i p_j q_{j|\{1,2\}}$, if node $i$'s battery is non-empty, we obtain
\begin{align}
\mu_{j|\mathrm{active}} & = p_j q_{j|\{j\}} \left[1-\min\left(\frac{\delta_i}{p_i}, 1 \right) \right] + [p_j (1-p_i) q_{j |\{j\}} + p_i p_j q_{j|\{1,2\}}]\min\left(\frac{\delta_i}{p_i}, 1 \right) \nonumber \\
& = p_j (q_{j |\{j\}} - \Delta_j \min(\delta_i, p_i))
\end{align}
Since the expected number of packets arriving into the data queue per slot is $\lambda_j$, it follows from the property of a stable system, that the input rate is equal to the output rate so that the probability that node $j$ is active at any given time slot is given by
\begin{equation}\label{eqn:prob_active}
\mathrm{Pr} [B_j \neq 0, Q_j \neq 0] = \frac{\lambda_j}{\mu_{j|\mathrm{active}}}
\end{equation}
which does not depend on $\delta_j$. As expectedly, however, if $\delta_j$ is decreased, the non-active slots will be more likely to occur due to the emptiness of the battery, rather than that of the queue. Nevertheless, what it says is that the ratio between the active and non-active slots must remain the same for a given input rate $\lambda_j$ as long as the queue is stable. By noting that node $j$ transmits with probability $p_j$ only when it is active and node $i$ is unstable, the limiting expectation on the actual rate serviced out of queue $i$ is obtained by 
\begin{align}
\lim_{n \rightarrow \infty} E[\mu_i(n)] & = \left\{ p_i q_{i|\{i\}} \left(1- \frac{\lambda_j}{\mu_{j|\mathrm{active}}} \right) + [p_i (1-p_j) q_{i |\{i\}} + p_i p_j q_{i|\{1,2\}}] \frac{\lambda_j}{\mu_{j|\mathrm{active}}} \right\} \min\left(\frac{\delta_i}{p_i}, 1 \right) \nonumber \\
& = \min(\delta_i, p_i) \left( q_{i |\{i\}} - \frac{\Delta_i \lambda_j}{q_{j|\{j\}} - \Delta_j \min(\delta_i, p_i)} \right) \nonumber \\
& \geqa \min(\delta_i, p_i) (q_{i |\{i\}} - \Delta_i \min(\delta_j, p_j)) \nonumber \\
& = \mu_i^s
\end{align}
where, for (a), we replaced $\lambda_j$ with $\mu_j^s - \epsilon_j$ and used the fact that $\epsilon_j$ is non-negative. Again, the instability of the queue $i$ implies that the LHS of \eqref{eqn:limit_queue_dynamics} is strictly positive and, then, it must be the case that 
\begin{equation}
\lambda_i = \mu_i^s - \epsilon_i > \lim_{n \rightarrow \infty} E[\mu_i(n)] \geq \mu_i^s
\end{equation}
which is impossible because $\epsilon_i$ is non-negative.

\end{proof}
\end{lemma}

\subsection{Necessary Condition}\label{sec:sec:necessary_condition}

The necessary condition for the stability of the considered system is derived through the construction of a hypothetical system; this hypothetical system operates as follows: i) the packet and energy chunk arrivals at each node occur at \textit{exactly} the same instants as in the original system, ii) the \textit{coin tosses} that determine transmission attempts at each node have \textit{exactly} the same outcomes in both systems, iii) however, one of the nodes in the system continues to transmit dummy packets even when its data queue is empty but its battery is non-empty. The dummy packet transmission continues to consume one chunk of energy in the battery but does not contribute to throughput if the transmission is successful. Such a construction of a hypothetical system with dummy packet transmissions has been widely used to analyze systems of interacting queues and yields a sufficient and necessary condition for the stability \cite{rao:stability, Szpankowski:stability, naware:stability, jeon:effect, pappas:optimal, pappas:wireless}. It uses the \textit{stochastic dominance technique}; the queue sizes in the new system are path-wise never smaller than their counterparts in the original system, provided the queues start with identical initial conditions in both systems. However, in the case of a system with batteries, as is considered in this work, there exist sample-paths on which this strict path-wise dominance is violated. This is because dummy packet transmissions alter the dynamics of the batteries through unproductive use of their contents. For example, there are instants when a node is no more able to transmit in the hypothetical system due to the lack of energy while it is able to transmit in the original system. Being not able to transmit may imply a better chance of success for the other node, if the latter attempts to transmit at those instants, which causes a collapse of the sample-path dominance. Instead, here we use the hypothetical system of transmitting dummy packets only to derive a \textit{necessary} condition for the stability of the original system. 

Let us define
\begin{equation}\label{eqn:sr_necessary}
    \mathcal{R}_i=\left\{ \boldsymbol{\lambda}: \lambda_i \leq \min(\delta_i, p_i) \left(q_{i|\{i\}} - \frac{\Delta_i \lambda_j}{q_{j|\{j\}} - \Delta_j \min(\delta_i, p_i)} \right), \lambda_j \leq \min(\delta_j, p_j) (q_{j|\{j\}} - \Delta_j \min(\delta_i, p_i)) \right\}
\end{equation}
where $i \neq j$ and $i, j \in \set{1,2}$.

\begin{lemma}\label{lemma:stability_necessary_condition}
If the system is stable under the slotted ALOHA, then $\boldsymbol{\lambda} \in \bigcup_{i \in \{1,2\}} \mathcal{R}_i$.

\begin{proof}
Let us consider a hypothetical system in which node $i$ transmits dummy packets when its packet queue is empty and node $j(\neq i)$ operates as in the original system, where $i,j \in \set{1,2}$. As mentioned earlier, all other random events including the packet arrivals, energy harvesting, and the decisions for transmissions have the same realizations as in the original system. In the hypothetical system, node $i$ transmits with probability $p_i$ regardless of the emptiness of its data queue, provided its energy queue is non-empty, and each transmission consumes one chunk of energy. Therefore, $B_i(n)$ forms a decoupled discrete-time $M$/$M$/1 queue whose probability of non-emptiness is given by $\min(\frac{\delta_i}{p_i}, 1)$ and node $i$ behaves independently from node $j$, i.e., node $i$ only has the effect of lowering the success probability of node $j$ in the average sense. The saturated throughput of node $j$, therefore, can be computed separately as in \eqref{eqn:saturated_throughput_two_queues} and the queue at node $j$ is stable if 
\begin{equation}\label{eqn:stability_condition1_dummy}
\lambda_j \leq \min(\delta_j, p_j) (q_{j|\{j\}} - \Delta_j \min(\delta_i, p_i))
\end{equation}
which follows by applying Lemma \ref{lemma:stability_sufficient_condition} to a single-node case. For $\lambda_j$ satisfying \eqref{eqn:stability_condition1_dummy}, the probability that node $j$ is active is obtained as in \eqref{eqn:prob_active} by noting that the probability that battery $i$ is non-empty is given by $\min(\frac{\delta_i}{p_i}, 1)$. Thus, the queue at node $i$ is stable if  
\begin{align}\label{eqn:stability_condition2_dummy}
\lambda_i & \leq \left\{ p_i q_{i|\{i\}} \left(1- \frac{\lambda_j}{\mu_{j|\mathrm{active}}} \right) + [p_i (1-p_j) q_{i |\{i\}} + p_i p_j q_{i|\{1,2\}}] \frac{\lambda_j}{\mu_{j|\mathrm{active}}} \right\} \min\left(\frac{\delta_i}{p_i}, 1 \right) \nonumber \\
& = \min(\delta_i, p_i) \left( q_{i |\{i\}} - \frac{\Delta_i \lambda_j}{q_{j|\{j\}} - \Delta_j \min(\delta_i, p_i)} \right) 
\end{align}
The pair of equations \eqref{eqn:stability_condition1_dummy} and \eqref{eqn:stability_condition2_dummy} describes the stability condition for the hypothetical system in which node $i$ transmits dummy packets, which is a necessary condition for the stability of the original system for the range of values of $\lambda_j$ specified in Eq. \eqref{eqn:stability_condition1_dummy}. The reason is this: if for some $\lambda_i$, queue $i$ is unstable in the hypothetical system, i.e., \eqref{eqn:stability_condition2_dummy} does not hold, then $Q_i(n)$ approaches infinity almost surely. Note that as long as queue $i$ does not empty, the behavior of the hypothetical system and the original system are identical, provided they start from the same initial conditions, since dummy packets will never have to be used. A sample-path that goes to infinity without visiting the empty state, which is a feasible one for a queue that is unstable, will be identical for both the hypothetical and the original systems. Therefore, the instability of the hypothetical system implies the instability of the original system.
\end{proof}
\end{lemma}

\subsection{Proof of Theorem \ref{thm:main_theorem}}\label{sec:sec:proof_main_theorem}

Here we first compute the closure of the outer bound of $\mathfrak{S}(\boldsymbol{\delta}, \boldsymbol{p})$ over all feasible transmission probability vectors $\boldsymbol{p} \in [0, 1]^2$. Therefore, any rate vector that is outside the closure is not attainable. After that, it is proven that the entire interior of the closure can be achieved by showing that the closure of the inner bound is identical with that of the outer bound.

Note that the description on the outer bound of $\mathfrak{S}(\boldsymbol{\delta}, \boldsymbol{p})$ in Lemma \ref{lemma:stability_necessary_condition} does not depend on $\boldsymbol{\delta}$ for $\boldsymbol{p} \preceq \boldsymbol{\delta}$ and also note that increasing $p_i$ over $\delta_i$ has no effect since the value of $\min(\delta_i, p_i)$ is bounded below by $\delta_i$. For the subregion $\mathcal{R}_i$, $i \in \set{1,2}$, let us consider the following boundary optimization problem in which we maximize the boundary of $\lambda_i$, denoted by $\bar{\lambda}_i$, for a given value of $\lambda_j$ $(j \neq i)$ as $\boldsymbol{p}$ varies\footnote{Note that optimizing the boundary of a region over $\boldsymbol{p}$ is equivalent to take the closure of the region over $\boldsymbol{p}$.}, that is
\begin{align}
\displaystyle \max_{\boldsymbol{p}} \ & \ \bar{\lambda}_i = p_i \left( q_{i|\{i\}} - \frac{\Delta_i \lambda_j}{q_{j|\{j\}} - \Delta_j p_i} \right) \label{eqn:opt_p1_obj}\\
\displaystyle \textrm{subject to} \ & \  \lambda_j \leq p_j (q_{j|\{j\}} - \Delta_j p_i) \label{eqn:opt_p1_const_1}\\
\displaystyle \ & p_i \leq \delta_i, \forall i \in \set{1,2} \label{eqn:opt_p1_const_2}
\end{align}
To maximize $\bar{\lambda}_i$ over $\boldsymbol{p}$, we need to understand their relationship. Note that $\bar{\lambda}_i$ depends only on $p_i$. Differentiating $\bar{\lambda}_i$ with respect to $p_i$ gives
\begin{equation}\label{eqn:y_first_derivative}
\frac{\partial \bar{\lambda}_i}{\partial p_i} = q_{i|\{i\}} - \frac{\Delta_i q_{j|\{j\}} \lambda_j}{(q_{j|\{j\}} - \Delta_j p_i)^2}
\end{equation}
and by differentiating once again, we have
\begin{equation}
\frac{\partial^2 \bar{\lambda}_i}{\partial p_i^2} = - \frac{ 2 \Delta_i \Delta_j q_{j|\{j\}} \lambda_j}{( q_{j|\{j\}} - \Delta_j p_i)^3}
\end{equation}
Since $q_{j|\{j\}} > \Delta_j$, the second derivative is negative and, thus, $\bar{\lambda}_i$ is a concave function of $p_i$. Equating the first derivative to zero gives the maximizing $p_i^{\ast}$ as
\begin{equation}\label{eqn:opt_prob}
p_i^{\ast} = \frac{1}{\Delta_j} \left( q_{j|\{j\}} - \sqrt{\frac{\Delta_i q_{j|\{j\}} \lambda_j}{q_{i|\{i\}}}} \right)
\end{equation}
and the corresponding maximum function value is obtained by substituting \eqref{eqn:opt_prob} into \eqref{eqn:opt_p1_obj}, thus yielding
\begin{equation}\label{eqn:maximum_y_curve}
{\bar{\lambda}}^{\ast}_{i,\mathrm{curve}} = \left( 1 - \sqrt{\frac{\Delta_i \lambda_j }{q_{i|\{i\}} q_{j|\{j\}}}} \right) \left(\frac{q_{i|\{i\}} q_{j|\{j\}} - \sqrt{\Delta_i q_{i|\{i\}} q_{j|\{j\}}} \lambda_j }{\Delta_j} \right)
\end{equation}
Suppose now that the maximum occurs at a strictly interior point of the feasible region, i.e., $p_i^{\ast} \in (0, \delta_i)$, which corresponds to the condition
\begin{equation}\label{eqn:range_1}
\frac{q_{i|\{i\}} (q_{j|\{j\}} - \Delta_j \delta_i)^2}{\Delta_i q_{j|\{j\}}} < \lambda_j < \frac{q_{i|\{i\}}q_{j|\{j\}}}{\Delta_i}
\end{equation}
which is obtained by rearranging Eq. \eqref{eqn:opt_prob} and substituting the extreme values of $p_i^{\ast}$.
On the other hand, the constraint \eqref{eqn:opt_p1_const_1} should also be satisfied for the derived $p_i^{\ast}$. Hence, by substituting \eqref{eqn:opt_prob} into \eqref{eqn:opt_p1_const_1} and using $p_j \leq \delta_j$, we obtain
\begin{equation}\label{eqn:range_2}
\lambda_j \leq \frac{\Delta_i q_{j|\{j\}} \delta_j^2}{q_{i|\{i\}}}
\end{equation}
Consequently, ${\bar{\lambda}}^{\ast}_{i,\mathrm{curve}}$ in \eqref{eqn:maximum_y_curve} is valid only for the range of values of $\lambda_j$ that satisfy both \eqref{eqn:range_1} and \eqref{eqn:range_2}. The intersection of the ranges of values of $\lambda_j$ determined by Eqs. \eqref{eqn:range_1} and \eqref{eqn:range_2} would be identical with the range specified by \eqref{eqn:range_1} if $\delta_j \geq \frac{q_{i|\{i\}}}{\Delta_i}$, which is impossible because $q_{i|\{i\}} > \Delta_i$ while $\delta_j \leq 1$. Thus, if $\Psi \geq 1$, where $\Psi$ is defined in Section \ref{sec:main_result}, the intersection is given by
\begin{equation}
\frac{q_{i|\{i\}} (q_{j|\{j\}} - \Delta_j \delta_i)^2}{\Delta_i q_{j|\{j\}}}  < \lambda_j \leq \frac{\Delta_i q_{j|\{j\}} \delta_j^2}{q_{i|\{i\}}}
\end{equation}
Otherwise, if $\Psi < 1$, the intersection is an empty set.

Next suppose that either $p_i^{\ast} = 0$ or $p_i^{\ast} = \delta_i$, which is the case when $\lambda_j$ lies outside of the range of Eq. \eqref{eqn:range_1}. If $\lambda_j$ is on the right-hand side of the range, i.e., if $\lambda_j \geq \frac{q_{i|\{i\}}q_{j|\{j\}}}{\Delta_i} $, $\bar{\lambda}_i$ is a non-increasing function of $p_i$ since its first derivative in \eqref{eqn:y_first_derivative} is non-positive. Therefore, $p_i^{\ast} = 0$ and $\bar{\lambda}_i^{\ast} = 0$. On the other hand, if $\lambda_j$ is on the left-hand side of the range of \eqref{eqn:range_1}, $\bar{\lambda}_i$ is a non-decreasing function of $p_i$ and, hence, $p_i^{\ast} = \delta_i$ and the corresponding maximum function value is obtained as
\begin{equation}\label{eqn:maximum_y_line}
{\bar{\lambda}}^{\ast}_{i,\mathrm{line}} = \delta_i \left( q_{i|\{i\}} - \frac{\Delta_i \lambda_j }{q_{j|\{j\}} - \Delta_j \delta_i} \right)
\end{equation}
for
\begin{equation}\label{eqn:range_line_1}
\lambda_j \leq \frac{q_{i|\{i\}} (q_{j|\{j\}} - \Delta_j \delta_i)^2}{\Delta_i q_{j|\{j\}}}
\end{equation}
On the other hand, the constraint \eqref{eqn:opt_p1_const_1} at $p_i^{\ast} = \delta_i$ becomes 
\begin{equation}\label{eqn:range_line_2}
\lambda_j \leq \delta_j (q_{j|\{j\}} - \Delta_j \delta_i)
\end{equation}
Thus, ${\bar{\lambda}}^{\ast}_{i,\mathrm{line}}$ is valid for the range of values of $\lambda_j$ specified as the intersection of the ranges given by \eqref{eqn:range_line_1} and \eqref{eqn:range_line_2}. If $\Psi \geq 1$, the intersection coincides with \eqref{eqn:range_line_1} and, if $\Psi < 1$, it coincides with \eqref{eqn:range_line_2}. To sum up, $\bar{\lambda}_i^{\ast}$ is obtained as follows:
\begin{itemize}
\item If $\Psi \geq 1$,
\begin{equation}
\bar{\lambda}_i^{\ast} = \left\{ \begin{array}{cl}
                \displaystyle {\bar{\lambda}}^{\ast}_{i,\mathrm{curve}}, & \ \mathrm{for} \ \frac{q_{i|\{i\}} (q_{j|\{j\}} - \Delta_j \delta_i)^2}{\Delta_i q_{j|\{j\}}}  < \lambda_j \leq \frac{\Delta_i q_{j|\{j\}} \delta_j^2}{q_{i|\{i\}}} \vspace{0.2cm}\\
                \displaystyle {\bar{\lambda}}^{\ast}_{i,\mathrm{line}}, & \ \mathrm{for} \ \lambda_j \leq \frac{q_{i|\{i\}} (q_{j|\{j\}} - \Delta_j \delta_i)^2}{\Delta_i q_{j|\{j\}}}
                \end{array}\right.
\end{equation}
\item If $\Psi < 1$,
\begin{equation}
\bar{\lambda}_i^{\ast} = {\bar{\lambda}}^{\ast}_{i,\mathrm{line}},  \ \ \mathrm{for} \ \lambda_j \leq \delta_j (q_{j|\{j\}} - \Delta_j \delta_i)
\end{equation}
\end{itemize}
Substituting $i \in \set{1,2}$ into the above yields the description for $\bar{\lambda}_1^{\ast}$ and $\bar{\lambda}_2^{\ast}$ which lead us to the description for the stability region given in Theorem \ref{thm:main_theorem}. Specifically, when $\Psi \geq 1$, the end points of ${\bar{\lambda}}^{\ast}_{2,\mathrm{curve}}$ and ${\bar{\lambda}}^{\ast}_{2,\mathrm{line}}$ and those of ${\bar{\lambda}}^{\ast}_{1,\mathrm{curve}}$ and ${\bar{\lambda}}^{\ast}_{1,\mathrm{line}}$ meet at $P_{B_1}$ and $P_{B_2}$ (which are defined in Section \ref{sec:main_result}), respectively. Furthermore, ${\bar{\lambda}}^{\ast}_{1,\mathrm{curve}}$ and ${\bar{\lambda}}^{\ast}_{2,\mathrm{curve}}$ are functions that are inverse of each other and they can be identically rearranged to coincide with \eqref{eqn:main_theorem}. These segments together with the axes form a closed region in the two-dimensional Euclidean space as shown in Fig. \ref{fig:sr_case1}. If $\Psi <1$, the end points of ${\bar{\lambda}}^{\ast}_{1,\mathrm{line}}$ and ${\bar{\lambda}}^{\ast}_{2,\mathrm{line}}$ meet at $P_{B_3}$ and, likewise, they define a closed region as shown in Fig. \ref{fig:sr_case2}. 

What is left to be shown is the achievability of the specified region. From Lemma \ref{lemma:stability_sufficient_condition}, we know that $\boldsymbol{\lambda} \preceq \boldsymbol{\mu}^s$ can be stably supported. For some $j \in \set{1,2}$, if $\boldsymbol{p} \preceq \boldsymbol{\delta}$, $\mu_j^s$ is written as
\begin{equation}
\mu_j^s = p_j (q_{j|\{j\}} - \Delta_j p_i)
\end{equation}
from which we derive
\begin{equation}\label{eqn:control_saturated_throughput}
p_j  = \frac{\mu_j^s}{q_{j|\{j\}} - \Delta_j p_i}
\end{equation}
By substituting $p_j$ into the expression for $\mu_i^s$ $(i \neq j)$, we have
\begin{equation}
\mu_i^s =  p_i \left( q_{i|\{i\}} - \frac{\Delta_i \mu_j^s}{q_{j|\{j\}} - \Delta_j p_i} \right) 
\end{equation}
which turns out to be identical to the expression for the outer boundary in \eqref{eqn:opt_p1_obj} by replacing $\mu_i^s$ with $\bar{\lambda}_i$ and $\mu_j^s$ with $\lambda_j$. In other words, the operating point of the saturated system can be controlled to any point on the boundary of $\mathcal{R}_i$ by adjusting $p_j$ according to \eqref{eqn:control_saturated_throughput} for $i \in \set{1,2}$ and $j \neq i$. This implies that the outer bound described by Lemma \ref{lemma:stability_necessary_condition} can be indeed achieved, which proves the achievability of Theorem \ref{thm:main_theorem}.

\section{The Impact of Finite Capacity Batteries}\label{sec:finite}

In this section, we consider the case where the capacity of the batteries is finite and study the impact of that on the previously obtained stability region. Denote by $c_i$ the capacity of the battery at node $i$. Then, the number of energy chunks stored in the battery evolves according to
\begin{equation}\label{eqn:battery_dynamics}
B_i(n+1) = \min \left( B_i(n) - 1_i(n) + H_i(n), c_i \right)
\end{equation}
i.e., the harvested energy chunks now can be stored only if the corresponding battery is not fully recharged. Since most of the analysis overlaps with the case of infinite capacity batteries, the result is demonstrated only for the collision channel model for brevity and to simplify the exposition. It becomes clear from the analysis that the channel with MPR capability can be handled similarly. Denote by $c_i$ the capacity of the battery at node $i$ and let $\lambda_i^{\max} \triangleq {\delta_i (1 - \delta_i^{c_i})}/({1 - \delta_i^{c_i+1}})$.

\begin{figure}[t]
\centering
\epsfig{file=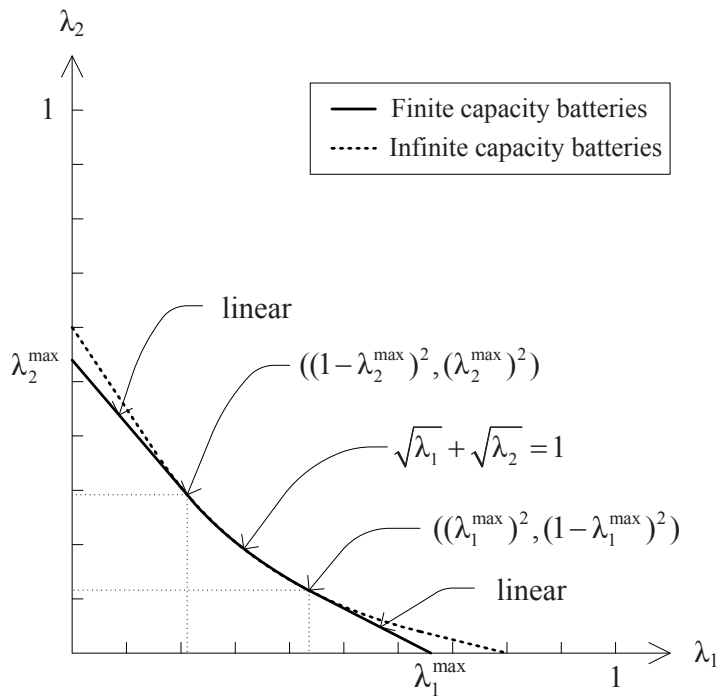,angle=0,width=0.52\textwidth}
\caption{Two-node stability region $\mathfrak{S}^{\boldsymbol{c}}(\boldsymbol{\delta})$ with finite capacity batteries where $\boldsymbol{\delta}=(0.8, 0.6)$ and $\boldsymbol{c}=(3,3)$ (the case when $\lambda_1^{\max} + \lambda_2^{\max} \geq 1$)}
\label{fig:stability_region_finite}
\vspace{0cm}
\end{figure}
\begin{theorem}\label{thm:finite}
For a given energy harvesting rate $\boldsymbol{\delta}$, the two-node stability region $\mathfrak{S}^{\boldsymbol{c}}(\boldsymbol{\delta})$ of the collision channel random access with batteries that have finite capacities, denoted by $\boldsymbol{c}=(c_i, i \in \set{1,2})$, is as follows. If $\lambda_1^{\max} + \lambda_2^{\max} \geq 1$, the boundary of the stability region is composed of three segments: (i) a line segment connecting $(0, \lambda_2^{\max})$ and $((1-\lambda_2^{\max})^2, (\lambda_2^{\max})^2)$, (ii) the curve $\sqrt{\lambda_1}+\sqrt{\lambda_2} = 1$ from the point with coordinates $((1-\lambda_2^{\max})^2, (\lambda_2^{\max})^2)$ to the point with coordinates $((\lambda_1^{\max})^2, (1-\lambda_1^{\max})^2)$, and (iii) a line segment connecting points $((\lambda_1^{\max})^2, (1-\lambda_1^{\max})^2)$ and $(\lambda_1^{\max}, 0)$. If $\lambda_1^{\max} + \lambda_2^{\max} <1$, the boundary is described by two straight line segments, namely, (i) the one connecting points $(0, \lambda_2^{\max})$ and $(\lambda_1^{\max}(1-\lambda_2^{\max}), \lambda_2^{\max}(1-\lambda_1^{\max}))$ and (ii) the one connecting $(\lambda_1^{\max}(1-\lambda_2^{\max}), \lambda_2^{\max}(1-\lambda_1^{\max}))$ and $(\lambda_1^{\max},0)$.
\begin{proof}
We begin by noting that when a node, say node $i$, transmits with probability $p_i$ whenever its battery is non-empty in case it is either saturated or transmits dummy packets if its queue is empty, the size of the battery, $B_i(n)$, follows a decoupled discrete-time $M$/$M$/1/$c_i$ model whose probability of being non-empty is given by \cite{kleinrock:queueing}
\begin{equation}\label{eqn:battery_nonempty_finite}
\begin{array}{lll}
    f_i \;=\; \left\{ \begin{array}{cl}
                \displaystyle \frac{\left(\delta_i/p_i\right) \left(1 - \left( \delta_i/p_i \right)^{c_i} \right)}{1 - \left(
                \delta_i/p_i
                \right)^{c_i+1}},\;& \mathrm{if} \ \delta_i \neq p_i \vspace{0.3cm}\\
                \displaystyle \frac{c_i}{c_i+1},\;& \mathrm{if} \ \delta_i = p_i
                \end{array}\right.
\end{array}
\end{equation}
By following similar steps as in the previous section, the outer bound of $\mathfrak{S}^{\boldsymbol{c}}(\boldsymbol{\delta}, \boldsymbol{p})$ for the case of the collision channel model is obtained as the union of regions described by
\begin{equation}\label{eqn:sr_outer_finite}
    \mathcal{R}_i=\left\{ \boldsymbol{\lambda}: \lambda_i \leq p_i f_i \left(1 - \frac{\lambda_j}{1 - p_i f_i} \right), \lambda_j \leq p_j f_j ( 1- p_i f_i) \right\}
\end{equation}
for $i \in \set{1,2}$ and $i \neq j$. Similarly, the saturated throughput vector of the system, which corresponds to an inner bound of the stability region, is obtained as $\boldsymbol{\mu}^s = (\mu_i^s, i \in \set{1,2})$ with $\mu_i^s = p_i f_i ( 1- p_j f_j)$. By substituting $p_j = \mu_j^s/ (f_j ( 1- p_i f_i))$ into the expression for $\mu_i^s$, we observe that the saturated throughput vector $\boldsymbol{\mu}^s$ can achieve any point on the boundary of $\mathcal{R}_i$, i.e., the rate region inside the outer bound of $\mathfrak{S}^{\boldsymbol{c}}(\boldsymbol{\delta}, \boldsymbol{p})$ can be indeed achieved.
Let us now compute the closure of the stability region. Observe that the function $p_i f_i$ is an increasing function of $p_i$ and, thus, its maximum occurs at $p_i^{\ast}=1$ and the corresponding maximum function value is obtained as ${\delta_i(1 - \delta_i^{c_i})}/({1 - \delta_i^{c_i+1}})$, which we denoted by $\lambda_i^{\max}$. Consequently, any non-negative function value that is less than or equal to $\lambda_i^{\max}$ can be attained by appropriately selecting a value for $p_i$ between zero and one. Based on this observation, replace $p_i f_i$ in \eqref{eqn:sr_outer_finite}
with another variable $p_i'\in [0, \lambda_i^{\max}]$. The rest of the proof follows that of Theorem \ref{thm:main_theorem}. Specifically, we set up and solve an optimization problem similar to that of Section \ref{sec:sec:proof_main_theorem} from \eqref{eqn:opt_p1_obj} to \eqref{eqn:opt_p1_const_2} with $q_{i|\{i\}} = 1$, $q_{i|\{1,2\}} = 0$, and $p_i \in [0, \lambda_i^{\max}]$, $\forall i \in \set{1,2}$. Indeed, the result for the channel with the MPR capability can be obtained by solving the same problem as in Section \ref{sec:sec:proof_main_theorem} but with $p_i \in [0, \lambda_i^{\max}]$, $\forall i \in \set{1,2}$. The remainder of the proof is omitted for brevity.
\end{proof}
\end{theorem}

An example two-node stability region $\mathfrak{S}^{\boldsymbol{c}}(\boldsymbol{\delta})$ with finite capacity batteries is depicted in Fig. \ref{fig:stability_region_finite} which is the region below the solid line. For comparison's sake, the stability region with infinite capacity batteries for the same parameter values is also depicted in the figure with the dotted line. The difference between the two regions, therefore, represents the loss due to the finite capacity of the batteries.

\begin{corollary}
Denote by $\mathfrak{S}^{\infty}(\boldsymbol{\delta})$ the stability region of the slotted ALOHA with infinite capacity batteries. Then, for any finite capacity batteries, the relation $\mathfrak{S}^{\boldsymbol{c}}(\boldsymbol{\delta}) \subset \mathfrak{S}^{\infty}(\boldsymbol{\delta})$ holds.
\end{corollary}
As observed in the proofs of Theorem \ref{thm:main_theorem} and Theorem \ref{thm:finite}, we can compute $\mathfrak{S}^{\infty}(\boldsymbol{\delta})$ and $\mathfrak{S}^{\boldsymbol{c}}(\boldsymbol{\delta})$ through a closure operation over $\boldsymbol{p}$ varying in the rectangles $[0, \delta_1] \times [0, \delta_2]$ and $[0, \lambda_1^{\max}] \times [0, \lambda_2^{\max}]$, respectively. The corollary follows from the fact that $\lambda_i^{\max}$ is strictly less than $\delta_i$ for any finite $c_i$.

\section{Concluding Remarks}\label{sec:conclusion}

We studied the effect of stochastic energy harvesting, which imposes energy availability constraint on each node, on the stability of the slotted ALOHA. An exact characterization of the stability region was carried out for the two-node case under the generalized wireless channel model with MPR capability. By comparing to the case of unlimited energy for transmissions, we identified the loss in terms of the size of the stability region for either infinite or finite battery capacities. To extend the analysis to more general networks presents serious difficulties of tractability due to the complex interaction between the nodes and may require approximations or alternative approaches that go beyond the scope of this paper.

\appendices

\section{On the Convexity of the Stability Region}\label{appendix:A}

In this section, the MPR probabilities are computed for a Rayleigh fading environment and the criterion on the convexity of the stability region given by $\Psi \gtrless 1$ is verified in terms of the values of physical layer parameters. The use of matched filters was implicitly assumed in using Eq. \eqref{eqn:sinr_criteria} for decoding the signal at the receiver, which fundamentally treats interference as white Gaussian noise. Although, techniques such as the successive interference cancellation \cite{tse:fundamentals} can improve the accuracy of the MPR capability description, comparing different physical layer techniques is outside the scope of our work here.

We begin by describing the SINR of the signal transmitted from node $i$ at the receiver as
\begin{equation}
\gamma_{i | \mathcal{M}} = \frac{P_{\mathrm{rx},i}}{N + \sum_{j \in \mathcal{M} \setminus \{i\}} P_{\mathrm{rx},j}}
\end{equation}
where $\mathcal{M}$ is the set of nodes transmitting simultaneously, $N$ is the background noise power, and $P_{\mathrm{rx},i}$ is the received signal power from node $i$ at the receiver which is modeled by
\begin{equation}
P_{\mathrm{rx},i} = \psi_{i}^2 K r_{i}^{-\nu} P_{\mathrm{tx},i}
\end{equation}
where $\psi_{i}$ is a Rayleigh random variable with $E[\psi_{i}^2]=1$, $K$ is a constant, $\nu$ is the propagation loss exponent, $r_{i}$ is the distance between node $i$ and the receiver, and $P_{\mathrm{tx},i}$ is the transmission power of node $i$. Let $f_{\psi_{i}^2}$ be the probability density function of $\psi_{i}^2$ which is exponential with unit mean \cite{goldsmith:wireless}. Then, the success probability of a transmission by node $i$ when it transmits alone is computed by
\begin{align}
q_{i|\{i\}} & = \mathrm{Pr}\left[ \frac{\psi_{i}^2 K r_{i}^{-\nu} P_{\mathrm{tx},i}}{N} \geq \theta \right] \\
& = \int_0^{\infty} \mathrm{Pr} \left[ \omega \geq \frac{\theta N r_{i}^{\nu}}{K P_{\mathrm{tx},i}} \right] f_{\psi_{i}^2}(\omega) d \omega \\
& = \exp\left( - \frac{\theta N r_{i}^{\nu}}{K P_{\mathrm{tx},i}} \right)
\end{align}
Similarly, the success probability of a transmission by node $i$ when it transmits along with the other node $j$ is given by
\begin{align}
q_{i|\{i, j\}} & = \mathrm{Pr} \left[ \frac{\psi_{i}^2 K r_{i}^{-\nu} P_{\mathrm{tx},i}}{N + \psi_{j}^2 K r_{j}^{-\nu} P_{\mathrm{tx},j} } \geq \theta \right] \\
& = \int_0^{\infty} \int_0^{\infty} \mathrm{Pr} \left[ \omega_i \geq \frac{ \theta (N+ \omega_j K r_{j}^{-\nu} P_{\mathrm{tx},j})}{K r_{i}^{-\nu} P_{\mathrm{tx},i}}\right] f_{\psi_{i}^2}(\omega_i)  f_{\psi_{j}^2}(\omega_j) d \omega_i d \omega_j \\
& = \int_0^{\infty} \exp \left( - \frac{ \theta (N+ \omega_j K r_{j}^{-\nu} P_{\mathrm{tx},j})}{K r_{i}^{-\nu} P_{\mathrm{tx},i}} \right) f_{\psi_{j}^2}(\omega_j) d \omega_j \\
& = \left( 1 + \theta \frac{P_{\mathrm{tx},j}}{P_{\mathrm{tx},i}} \left( \frac{r_{i}}{r_{j}} \right)^{\nu}\right)^{-1} \exp\left( - \frac{\theta N r_{i}^{\nu}}{K P_{\mathrm{tx},i}} \right)
\end{align}
where $i, j \in \{1,2\}$, $j \neq i$, and $\psi_{i}$ and $\psi_{j}$ were assumed mutually independent.

From Corollary \ref{cor:convexity}, we know that the convexity of the stability region $\mathfrak{S}(\boldsymbol{\delta})$ is determined by $\Psi \gtrless 1$ where $\Psi$ was defined as
\begin{equation}\label{eqn:appendix_equivalence}
\Psi = \frac{\Delta_1 \delta_2}{q_{1|\{1\}}} + \frac{\Delta_2 \delta_1}{q_{2|\{2\}}} 
\end{equation}
By substituting the obtained MPR probabilities above into the definition of $\Delta_i$, we obtain
\begin{equation}
\Delta_i = \theta \frac{P_{\mathrm{tx},j}}{P_{\mathrm{tx},i}} \left( \frac{r_{i}}{r_{j}} \right)^{\nu} \left( 1 + \theta \frac{P_{\mathrm{tx},j}}{P_{\mathrm{tx},i}} \left( \frac{r_{i}}{r_{j}} \right)^{\nu}\right)^{-1} \exp\left( - \frac{\theta N r_{i}^{\nu}}{K P_{\mathrm{tx},i}} \right)
\end{equation}
Then, by substituting $\Delta_i$, $\forall i \in \{1,2\}$, into \eqref{eqn:appendix_equivalence}, we express $\Psi$ in terms of the physical layer parameters as 
%
\begin{equation}\label{eqn:appendix_convexity}
\Psi = \frac{\theta P_{\mathrm{tx},2} r_{1}^{\nu} \delta_2}{P_{\mathrm{tx},1} r_{2}^{\nu} + \theta P_{\mathrm{tx},2} r_{1}^{\nu}} + \frac{\theta P_{\mathrm{tx},1} r_{2}^{\nu} \delta_1}{P_{\mathrm{tx},2} r_{1}^{\nu} + \theta P_{\mathrm{tx},1} r_{2}^{\nu}} \gtrless 1 
\end{equation}
Interestingly, in the case of having unlimited energy for transmissions, i.e., $\delta_i = 1$, $\forall i \in \set{1,2}$, the above criterion \eqref{eqn:appendix_convexity} simplifies to $\theta \gtrless 1$, which does not depend on any other parameters than the threshold $\theta$.

\bibliographystyle{IEEEtran}
\bibliography{bib_all}

\end{document}